\documentclass[twocolumn,twoside,slac_two]{revtex4}
\usepackage{graphicx}
\usepackage{epstopdf}

\usepackage{fancyhdr}
\pdfoutput=1

\begin{document}

\title{Flaring gamma-ray AGNs seen with the \textit{FERMI} LAT}

\author{S.~Buson}
  \affiliation{Universit\`{a} degli Studi di Padova \& INFN Padova, via Marzolo 8, I-35131 Padova}
\author{D.~Bastieri}
\affiliation{Universit\`{a} degli Studi di Padova \& INFN Padova, via Marzolo 8, I-35131 Padova}
\author{F.~D'Ammando}
\affiliation{Dipartimento di Fisica \& INFN Perugia,  Via A. Pascoli I-06123, Perugia}
\author{G.~Tosti}
\affiliation{Dipartimento di Fisica \& INFN Perugia,  Via A. Pascoli I-06123, Perugia}
\author{on behalf of the Fermi-LAT Collaboration}
\noaffiliation{}

\begin{abstract}
In the first 3.5 years of operations \textit{Fermi} detected several sources 
with daily fluxes brighter than $F(E>100\, \mathrm{MeV}) > 10^{-6}\; \mathrm{ph}\ \mathrm{cm}^{-2}\ \mathrm{s}^{-1}$, the threshold set by the \textit{Fermi} Collaboration for issuing an Astronomer Telegram (ATel).
We focus the attention on these flaring sources, most of which are blazars,  known to be extremely variable over the whole electromagnetic spectrum, from radio to gamma-ray energies and we investigate
 the properties of the selected sample  comparing them to the general characteristics of the \textit{Fermi} catalog sources.

\end{abstract}

\maketitle

\thispagestyle{fancy}


\section{INTRODUCTION}
Since the beginning of its operations the Large Area Telescope (LAT) on-board the \textit{Fermi} satellite has guaranteed a continuous monitoring of the MeV-GeV sky.
 Notably, the usual LAT operation mode, the scanning sky-survey mode, allows the coverage of the full sky every 
 two orbits (about 3 hours), therefore offering a great opportunity for rapid flare detection and follow-up observations. 
 Complement to this  all-sky monitoring capability of variability and transients detection is the
invaluable effort offered by  the Flare Advocate duty activity (also known as Gamma-ray Sky Watcher, FA-GSW), a scientific service belonging 
 to the LAT Instrument Science Operations aiming to provide a first quick-look inspection and daily review of the 
 gamma-ray sky observed by the LAT, carried out with continuity all the year round through weekly shifts.
Sources that are found to exceed the predefined flux daily scale of 
$F(E>\,100\,MeV) > 10^{-6}\ \mathrm{ph}\ \mathrm{cm}^{-2}\ \mathrm{s}^{-1}$
 (which is a typical value used for defining a source in flare at gamma rays) 
are promptly reported to the scientific community, e.g. by means of ATels.

In the period from August 4, 2008 to February 4, 2012,  91 of them fulfilled the ATel criterion as they underwent at least one flaring state overcoming the predefined flux threshold.
 We collect all these flaring events and
 we note that among them, the overwhelming of flaring objects is composed by blazars. 
 Actually, although they represent a relatively rare sub-class of Active Galactic Nuclei (AGNs, 10\% of the entire AGN population),  these objects constitute the largest known population of gamma-ray sources as well as the great majority of sources detected by the \textit{Fermi} LAT.
 We focus on the general characteristic of the sample by itself, paying particular attention to the general properties 
 of the AGNs pointed out by the FA-GSW activity and provide a comparison with the AGNs seen in flaring state by the previous gamma-ray instrument, the Energetic Gamma-Ray Experiment Telescope (EGRET), on-board the Compton Gamma-Ray Observatory \citep{Thompson93}.

\section{SAMPLE COMPOSITION}

The objects that exceeded the fixed flux threshold that have been reported in \textit{Fermi} Atels belong to different source classes.
Following the classification that has been used in the second \textit{Fermi} catalog of AGNs \citep[2LAC,][]{2lac}
the objects of this sample divide in: 68 Flat Spectrum Radio Quasars (FSRQs), 14 BL Lacs, 2 AGN, 4 AGU\footnote{AGU refers to sources without a good optical spectrum or without an optical spectrum at all, whereas AGN refers to sources that are not confirmed blazars nor blazar candidates. For more details we address the reader to \citet{2lac}.}, 
2 Narrow Line Seyfert I (NLSyI) and one Radio Galaxy.

Among them, 83 have a measured redshift and 7 have not been detected in the second \textit{Fermi} source catalog  \citep[2FGL,][]{nolan}. Considering the usual BL Lac classification based on the position of the SED peak,  the 14 sources are divided in 6 ISP, 6 LSP and 2 HSP \citep[as defined in][]{2lac}. 
Worth to mention is that nine sources of this sample are detected at TeV energies: Ap Lib, S5 0716$+$74, BL Lacertae, 3C 66A, PKS 2155$-$304, 3C 279, 4C $+$21.35, PKS 1510$-$089 and NGC 1275. 

\section{\textit{Fermi} LAT DATA ANALYSIS}

The \textit{Fermi} LAT is   a pair conversion telescope with large effective area ($\sim$\,8000 cm$^2$ on axis for E\,$>1$\,GeV) and large field of view ($\sim$\,2.4 sr at 1 GeV).  It is optimized for gamma
rays in the energy range from 20\,MeV up to energies beyond 300\,GeV. 
Full details of the instrument and descriptions of the on-board and ground data processing are provided in \citet{LAT09_instrument}, and information regarding on-orbit calibration procedures is given by \citet{LAT12__calib}. 
Thanks to its high sensitivity and almost uniform sky coverage it is an ideal tool for multiwavelength monitoring,  gamma-ray flares detection and follow-up observations.

For our analysis we select all  ``source'' class events located within $10^{\circ}$ of each source of interest for the time interval reported in the correspondent ATel (one day interval).                         
When a time interval larger than a day is indicated in the ATel, we look at the daily Automated Science Processing (ASP) light curve and analyze the data related to the first daily peak showed by the ASP light curve.
To limit the contamination from  Earth-limb gamma rays produced by the cosmic-rays interaction with the upper atmosphere, data are restricted to a maximal zenith angle of $100^{\circ}$ and time periods when the spacecraft rocking angle exceeded $52^{\circ}$ are excluded.
For each flaring episode we first use the {\it pointlike} tool to derive the best estimation on the position of the flaring source.
 Subsequently the analysis is performed with the standard analysis tool  {\it gtlike}, part  of the \textit{Fermi} ScienceTools software package (version 09-27-01) available from the \textit{Fermi} Science Support Center (FSSC\footnote{Web address: \texttt{http://fermi.gsfc.nasa.gov/ssc/}}). 
An unbinned maximum likelihood technique \citep{mattox96} is applied to events in the energy range from 100\,MeV to 300\,GeV  in combination with the post-launch instrument response functions (IRFs) P7SOURCE\_V6 to derive the spectral parameters. 
\\
The model for each Region of Interest (RoI) includes the diffuse Galactic foreground emission  and 
the isotropic diffuse emission: the former by means of the template gal\_2yearp7v6\_v0\_trim.fits and the latter, which accounts  for 
both photons and residual charged particle background, by means of the template iso\_p7v6source.txt\footnote{These templates are publicly  available at the web address: \texttt{http://fermi.gsfc.nasa.gov/ssc/data/access/lat/ \\BackgroundModels.html}}. 
Sources reported in the 2FGL catalog and located within $15^{\circ}$ of the target sources are incorporated in the model of the RoI as well. For these individual LAT sources and the sources of interest themselves we assumed a power-law model.
In the fit the parameters of sources located within $10^{\circ}$ radius centered on the source of interest are left free while parameters of sources located within the $10^{\circ}$-$15^{\circ}$ annulus are fixed and the isotropic normalization as well. 
We perform a first fit to estimate reliable  starting values for the parameters in the RoI and remove from the model sources with too low significance. For this aim we use the Test Statistic (TS) which provides an estimate of the significance of the detection for each gamma-ray source in RoI. The TS value is defined as twice the difference between the log-likelihood function maximized by adjusting all the parameters of the model, with and without the source. Sources with TS lower than 5 are removed from the RoI model.
A subsequent minimization procedure is applied to the data to derive the final values for the fluxes and spectral indices.
 When the target source is not significantly detected, i.e. the TS value is lower than 25\footnote{The square root of TS in the case of two degree of freedom is distributed as $\chi^2$, therefore TS=25 roughly corresponds to 4.6 sigma.}, flux upper limits at 95\% confidence level are calculated. 

\section{GENERAL CHARACTERISTICS}
All the recalculated values and the complete outcomes resulting from this analysis will be presented in an upcoming publication, 
in the following we briefly report on the general characteristics of the sample.
In particular we compare the properties of our sample with those of the whole sources detected by \textit{Fermi} considering the 2FGL and 2LAC catalogs.

\begin{figure}
\includegraphics[width=85mm]{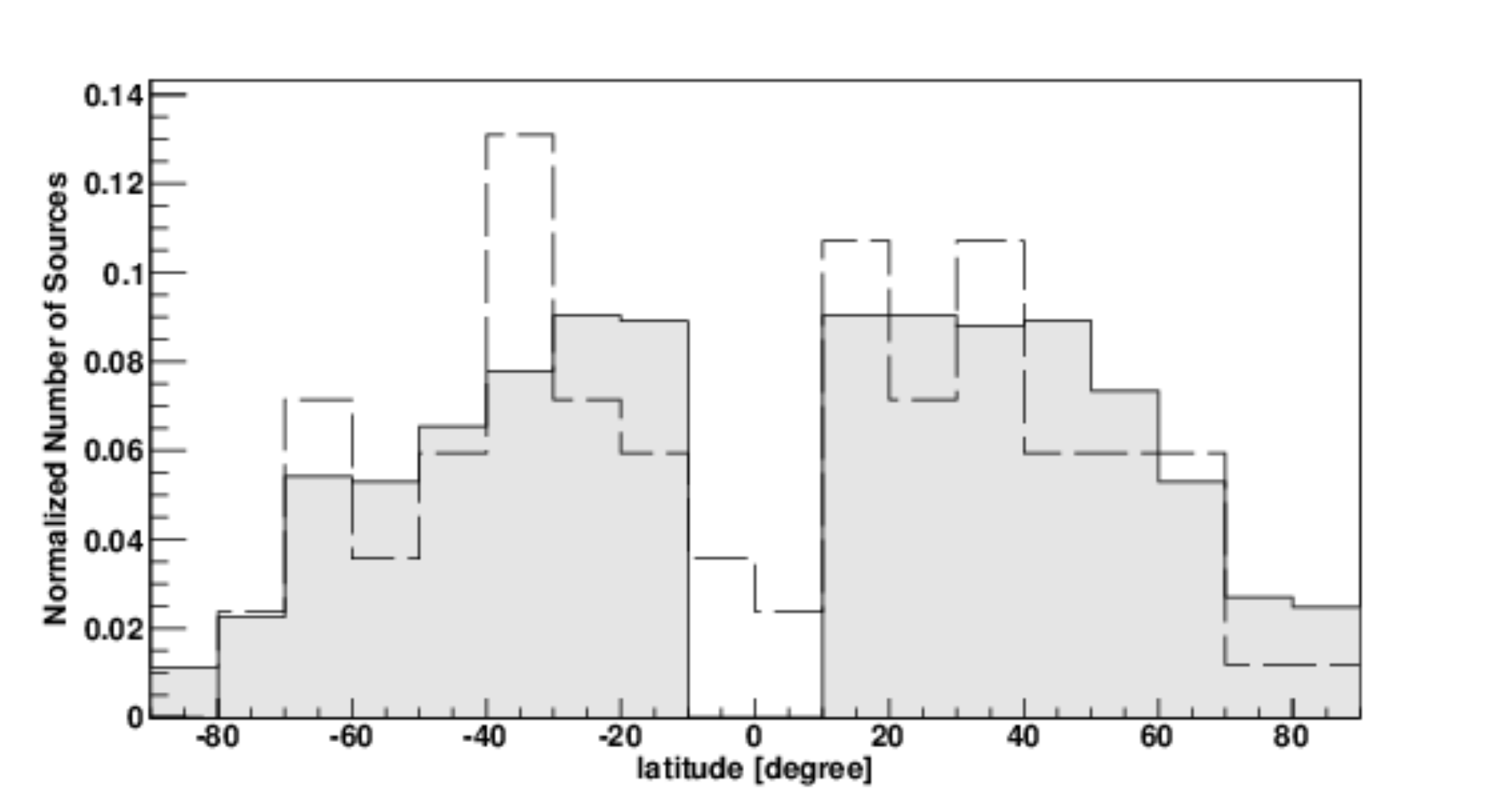}
\caption{Distribution of sources in Galactic latitude: solid line (gray area) represents the overall distribution corresponding to the 2LAC sample, dashed line this sample (normalized counts on vertical axis).}
\label{fig:lat}
\end{figure}

\begin{figure}
\includegraphics[width=85mm]{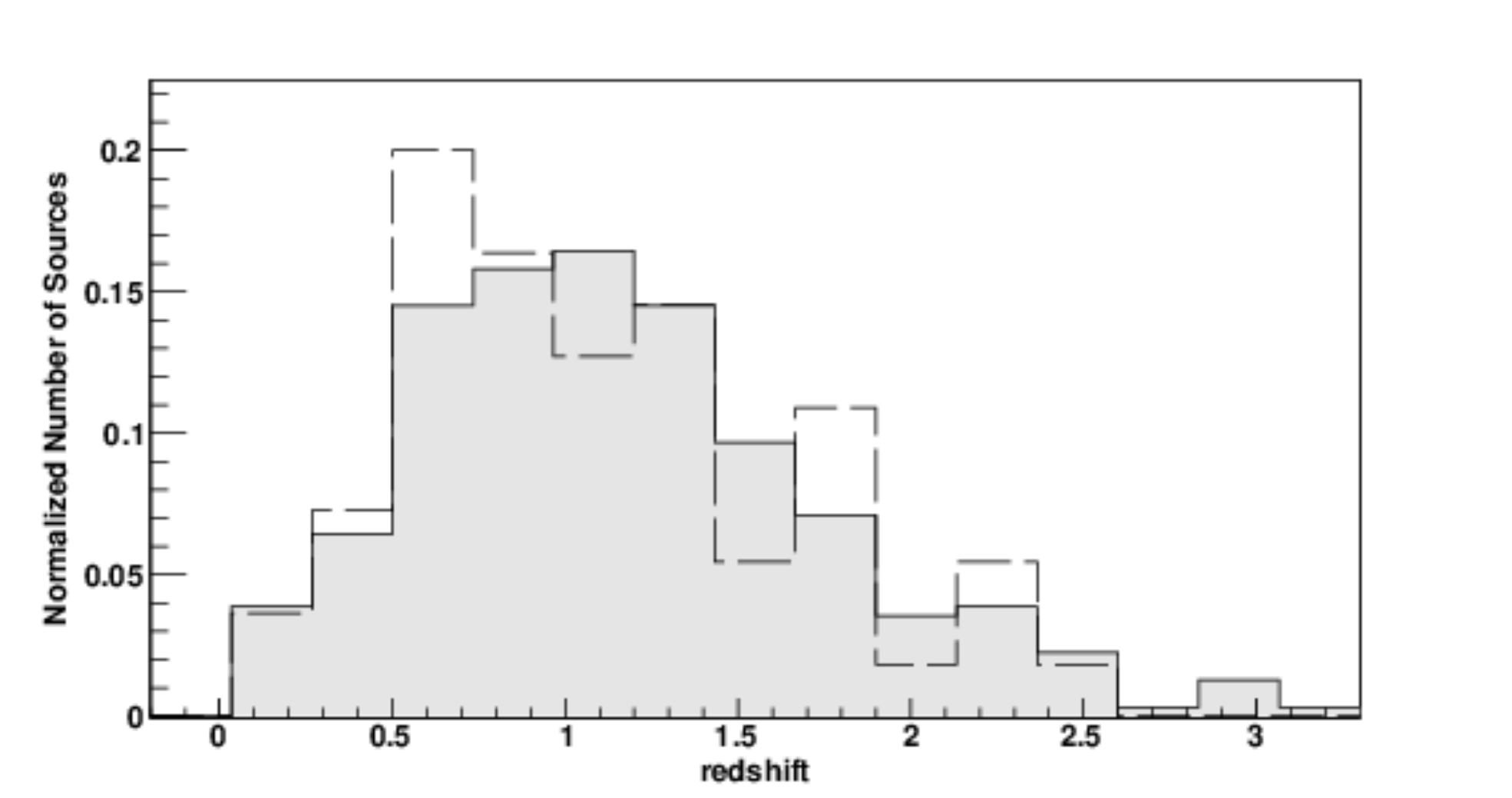}
\caption{Redshift distribution: solid line (gray area) represents the overall distribution corresponding to the FSRQ 2LAC clean sample, dashed line the FSRQ of this sample (normalized counts on vertical axis).}
\label{fig:z}
\end{figure}

\begin{figure}
\centering
\includegraphics[width=85mm]{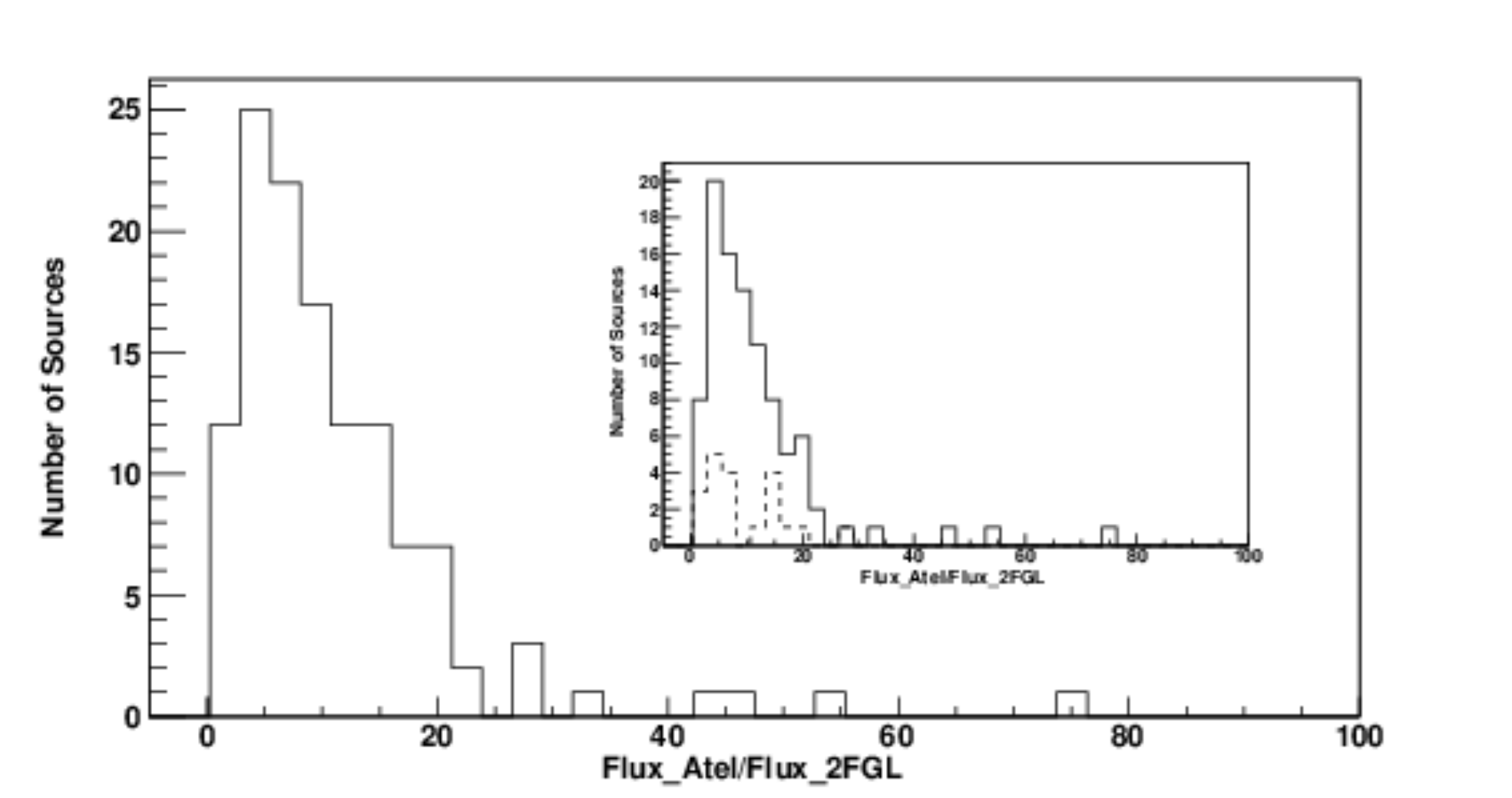}
\caption{Ratio between the flux (above 100\,MeV) measured in the time interval reported in the Atel and the mean 2FGL flux; in the inset are shown the two main different classes: FSRQ solid line, BL Lacs dashed line.}
\label{fig:f}
\end{figure}

\begin{figure}
\includegraphics[width=85mm]{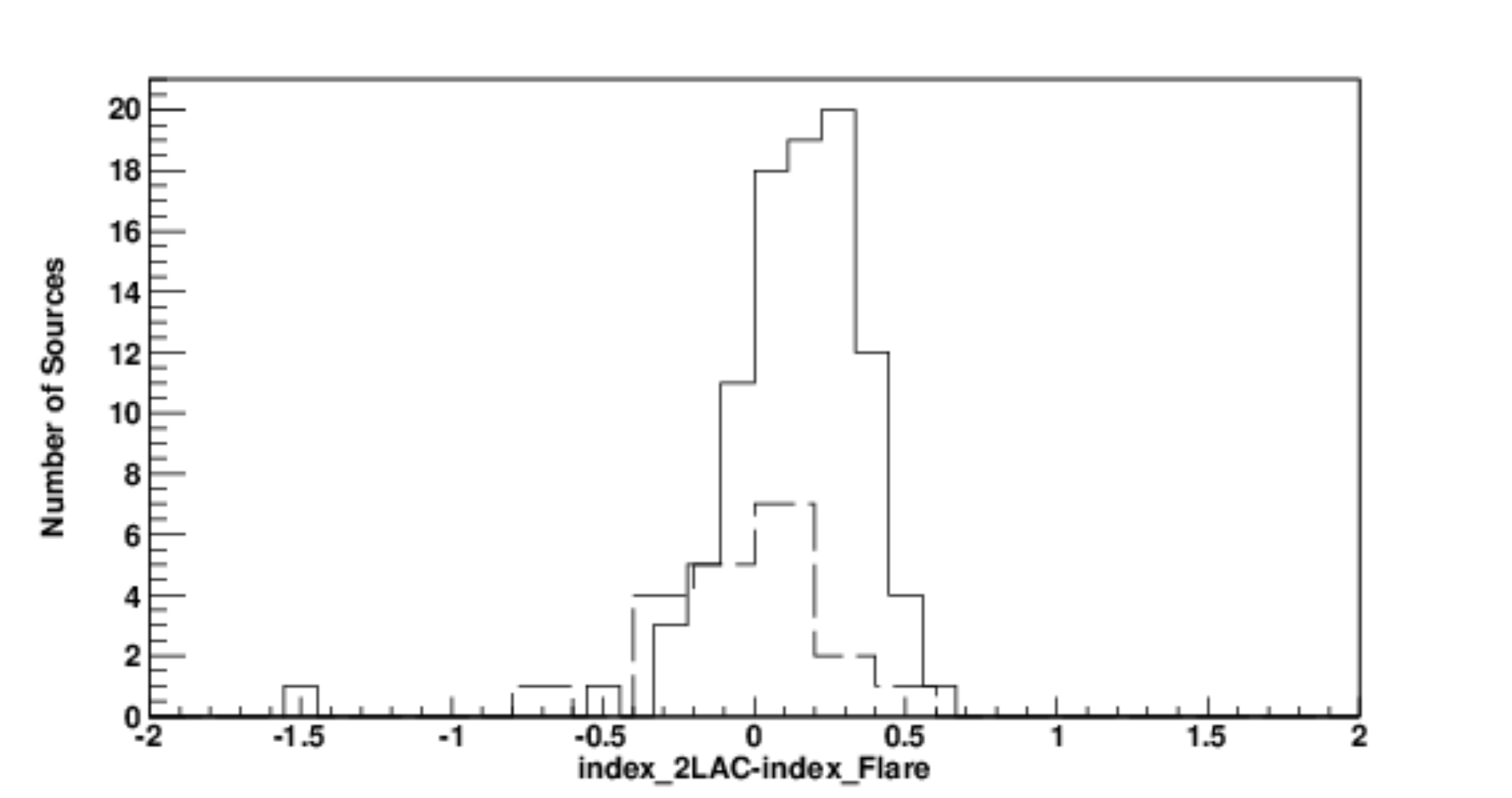}
\caption{Difference between the photon index reported in the 2LAC (which is computed over a period of 2 years)
and the index measured in the time interval reported in the ATel. FSRQ are plotted with solid line, BL Lacs with dashed line; here only sources belonging to the 2LAC clean sample are considered.}
\label{fig:ind}
\end{figure}

The sources of this sample  appear to be equally distributed in the northern and southern hemisphere as evidenced by
Fig.~\ref{fig:lat} which shows the histogram of the Galactic latitude distribution for the objects of this study (dashed line) and the ones reported in the 2LAC clean catalog (solid line, gray area). 
The only discrepancy  is found at low latitudes ($|b| <10^\circ$), 
due to the fact that  the 2LAC clean sample is build upon  
the a priori criterion to select only sources at high Galactic latitude, i.e.\@ $|b| > 10^\circ$.
The redshift distribution of this sample (dashed line) is presented in Fig.~\ref{fig:z} and follows well the redshift distribution of the 2LAC clean sample (solid line, gray area).
Fig.~\ref{fig:f}  shows the flux distribution of our sample as the ratio between the flux calculated for the time interval reported in the ATel and the average value reported in the 2LAC.
 We find that, on average, the sources of this sample  have fluxes a factor 11 higher than their 2LAC averages. 
 Despite this, there are some exceptional cases for which we measure a flaring flux $>40$ times its average value (as in the case of 2FGL J0532.0$-$4826, 2FGL J1153.2$+$4935 and 2FGL J1848.6$+$3241).
The flux distributions of the two main classes that make up our sample are plotted in the inset of Fig.~\ref{fig:f}: FSRQs are represented by the solid line while BL Lacs by the dashed line. These distributions do not indicate strong differences between the classes, although we note that the  
higher flux increments are displayed only by the former.
The distribution of the photon index derived for the flaring state is presented in Fig.~\ref{fig:ind}, 
evidencing the two classes of blazars: BL Lacs are plotted with a dashed line, and FSRQs with a solid line.
The histogram represents the difference between the average 2LAC index and the index measured during the flare.
While BL Lac indices remain almost constant, FSRQs show a hint of hardening  their spectra during the flaring state.

\section{EGRET COMPARISON}
Among the 271 pointlike sources detected in gamma rays by EGRET \citep[see e.g.][]{mukherjee01} 66 were blazars
plus a radio galaxy (Centaurus A), and some of them showed variability, being observed
several times at different amplitudes
\citep{hartman99}. From a comparison of the AGNs detected by EGRET with the objects
observed in flaring activity by {\it Fermi} LAT during the first 3.5 years of
operation, results that 29 of the 67 EGRET sources have been detected in high 
gamma-ray state by {\it Fermi}. Focusing on 
the gamma-ray sources detected by EGRET with a flux $F(E>100\, \mathrm{MeV}) > 10^{-6}\; \mathrm{ph}\ \mathrm{cm}^{-2}\ \mathrm{s}^{-1}$, 10 of the 16 flaring sources observed by EGRET
\citep[see e.g.][]{mukherjee01} have displayed intense gamma-ray activity also
during the {\it Fermi} operation.
This is an indication that the high activity
of these sources continued at least over two decades, even if in some cases the source
underwent a quiescent phase before increasing again its activity. A
typical example is PKS 0528$+$134, that was observed by EGRET during different
flaring episodes between 1991 and 1997 \citep{mukherjee09}, while it
remained quiescent during the {\it Fermi} observation up to a new flare in
June 2011  \citep{dammando11}. 
On the contrary, e.g.\@ 3C 279 was active for most
of the time during both the EGRET and {\it Fermi}  observations. 
However,
most of the flaring AGNs detected by the LAT did not show significant
gamma-ray activity during the EGRET observations. In this context, it is
indicative the behavior observed on 3C 454.3 and PKS 1510$-$089, the two
brightest extragalactic blazars in the {\em Fermi} sky.
During the EGRET
observations only a moderate activity was observed from these sources; on the contrary
they were characterized by an intense and dynamic behavior 
 during the \textit{Fermi} monitoring, reaching a remarkable  gamma-ray
flux higher than  $F(E>100\, \mathrm{MeV}) > 10^{-5}\; \mathrm{ph}\ \mathrm{cm}^{-2}\ \mathrm{s}^{-1}$ \citep{sanchez10,hungwe11}. 

Indeed, taking into consideration the different operating mode
of EGRET and LAT, i.e. inertial pointing versus all-sky scanning survey mode,
it cannot be excluded that some strong gamma-ray flares  have been missed for lack of observation coverage in the EGRET era.
Notwithstanding, the finding that some blazars detected by EGRET and not by {\em Fermi} suggests a possible significant change of activity in some objects, at least between the epochs of EGRET and {\em Fermi} operation.

\section{CONCLUSIONS}
During the first 3.5 years of \textit{Fermi} operations 91 sources displayed intense gamma-ray activity with fluxes exceeding 
$F(E>\,100\,\mathrm{MeV}) > 10^{-6}\  \mathrm{cm}^{-2}\ \mathrm{s}^{-1}$.
We studied this sample of sources  comparing their characteristics to the 2FGL and 2LAC samples and 
we showed that their properties are similar to the main {\it Fermi} LAT samples.
It is worth to notice that this sample constitutes only 10\% of all the sources detected in 2 years of \textit{Fermi} operations. Therefore, up to now, only a small percentage of the 2FGL sources displayed variability reaching extreme flux values, and those are basically all blazars.
All the flaring episodes have been analyzed in details and the correspondent results will be presented  in the upcoming publication where also further considerations will be addressed.

\bigskip 
\begin{acknowledgments}
The $Fermi$ LAT Collaboration acknowledges support from a number of agencies and institutes for both development and the operation of the LAT as well as scientific data analysis. These include NASA and DOE in the United States, CEA/Irfu and IN2P3/CNRS in France, ASI and INFN in Italy, MEXT, KEK, and JAXA in Japan, and the K.~A.~Wallenberg Foundation, the Swedish Research Council and the National Space Board in Sweden. Additional support from INAF in Italy and CNES in France for science analysis during the operations phase is also gratefully acknowledged.
\end{acknowledgments}


\end{document}